# Theoretic Insight into $CO_2$ Reduction at Active Sites of Molybdenum and Tungsten Enzymes: a π Interaction between $CO_2$ and Tungsten Bis-Dithiolene Complexes.

Yong Yan*†, Jing Gu †

Department of Chemistry, Princeton University, Princeton, NJ 08540

*Supporting Information Placeholder*

**ABSTRACT:** Active sites of molybdenum and tungsten enzymes, particularly mononuclear tungsten formate dehydrogenase (FDH) have been theoretically investigated towards their interaction with $CO_2$. Obvious π interaction has been found between the 2e reduced metallodithiole moiety and the molecular $CO_2$. This weak π bonding is predicated both at gas phase, noted as -6.0 kcal/mol and aqueous solvation level, -3.6 kcal/mol. Such interaction is not only limited to $CO_2$, but also to the $CO_2$ reduced product, i.e. formate, in the form of anion- π interaction, noted as -6.8 kcal/mol and -4.1 kcal/mol respectively in gas and aqueous solvation model. The Bailar twisted angles from 60° to 0°, governing structure preference of tungsten dithiolene from octahedron to triangle prism in their restricted structures, has been explored to evaluate such π interrelations with $CO_2$ and formate. An octahedral structure with 3 kcal/mol energy lower is preferred over the triangle prismatic when such interactions are concerned.

Recently, solar energy storage into chemical bonds, for example, dihydrogen, based on solar induced water splitting;[1] alcohols or other energy-dense carbon-based molecules from catalytic carbon dioxide reductions, has been of boundless interest.[2] Particularly, great attentions have been paid to later not only because the carbon-based high energy products or fuels exhibit more convenience in transportation than hydrogen, more compatible with current energy consumption system, but also due to the fact that this approach can mitigate anthropogenically produced $CO_2$, a greenhouse gas whose atmospheric concentrations have severely increased since industrial revolution.[2a]

Significant catalysis researches have been impelled with several approaches, including, homogenous catalytic reduction of $CO_2$ by organic molecules (i.e. pyridinium) or transition metal complexes (Re(bpy)Cl3, Ni(cyclam)2, Fe(porphyrin), etc.) in organic solvents (or organic solvents mixed with water);[3] heterogeneous photoelectrocatalytic reduction usually incorporated with semiconductor electrodes (i.e. GaP, GaAs); or the combination of these homogenous catalyst and semiconductor electrodes.[4] A recent publication further addressed such approach, learning from methanoptrin, to electrochemically reduce $CO_2$ in aqueous solution to form formic acid, formaldehyde and methanol in reasonable efficacy, solely via a simple pterin electrocatalyst, 6,7- dimethyl-4-hydroxy-2-mercaptopteridine (PTE).[5] Several other approaches are also conducted under the nature's guidance, for example, CODH enzyme (active center: molybdenum pterin monodithiolenes) for $CO_2$ conversion to CO, or similar dithiolene containing active site in formate dehydrogenase (FDH, active site: molybdenum or tungsten bis-dithiolenes) for fast electrocatalytic carbon dioxide reduction to form formate when FDH was attached on a graphite "pot" electrode.[6]

The extraordinary structural and bonding properties of molybdenum or tungsten containing dithiolene complexes have been extensively studied by us [7] and others.[6a,8] Particularly, the structure of a carbonyl ligated tungsten dithiolene can be inter-converted between trigonal prismatic and octahedral geometries. Such monodithiolene tetracarbonyl complex of tungsten's structural preference was found to be exclusively governed by the redox states of this complex.[7c] A series of ancillary ligands of tungsten bis-dithiolene has also been investigated to uncover the ligand's effect on the dithiolene's noninnocence

properties.[7e] However, such mononuclear molybdenum or tungsten dithiolene complex's investigation directly targeting carbon dioxide has been of fewer studies, with noticeable exception, i.e. Russo et al. have theoretically explored the reaction mechanism of molybdoenzyme formate dehydrogenase.[6b] This manuscript focused on a completely different approach: weak interaction of $CO_2$ or $CO_2$ reduced products with different redox states of tungsten bis-dithioene complex. Hence, we avoid the controversial pathway for $CO_2$'s insertion; while we exclusively probe the weak interaction that might be providing positive assistance for $CO_2$'s reaction with reduced states of tungsten dithiolenes and for reduced products, i.e. formate, to leave the mimic active site of tungsten.

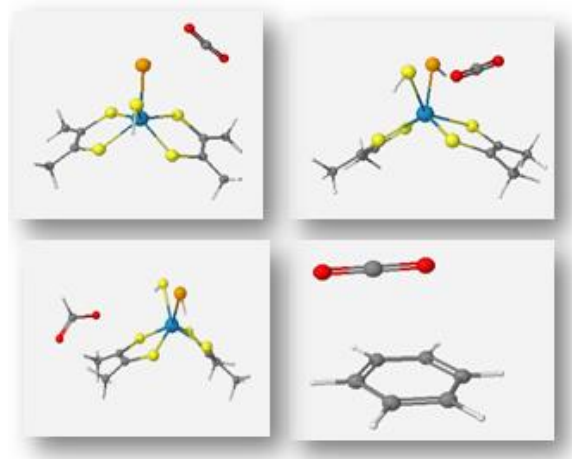

Figure 1. Optimized structures of $a^{2-}$-$CO_2$ (up left); $a$-$CO_2$ (up right); $a$-$HCO_2^-$ (bottom left); benzene-$CO_2$ (bottom right). Atom color: blue-tungsten; red-oxygen; yellow-sulfur; orange-selenium; gray-carbon; white-hydrogen.

Tungsten bis-dithiolene complex $a$ as show in Figure 1 was the target molecule; such simple model has been previously explored successfully to simulate the FDH.[6a,b] The theoretical approach was achieved based on density functional theory according to our previous published method as detailed in following:[7c,9,10] Geometry optimizations of these complexes were performed with the Gaussian 09 suite of software and employed the B3LYP functional. The authenticity of each converged structure was confirmed by the absence of imaginary vibrational frequencies. A double-ζ (DZ) basis set with an effective electron core potential (LANL2DZ ECP) was used for tungsten, a triple-ζ basis with two polarization functions was used for sulfur and selenium, and basis sets [(6-311+(d,p), 6-31(d,p), 6-31(d)] were all evaluated in comparison for the remaining atoms. Orbital images were created with the use of jmol. The ten-degree incremental Bailar twist rotations were performed by Chemcraft with intraligand bonds distances held constant for the dithiolene ligand. The polarizable continuum model (PCM) was applied to model solvent effects.

First, there basis sets [(6-311+(d,p), 6-31(d,p), 6-31(d)] for C, H, O were evaluated for $CO_2$'s interaction with complex $a$ and its two reduced states $a^{2-}$. A very clear π interaction was observed in the converged structure as shown in Figure 1 and as indicated by negative energy of such π interaction $\Pi(a^{2-}$-$CO_2)$, as defined:

$$\Pi (a^{2-}\text{-}CO_2) = E(a^{2-}\text{-}CO_2) - E(a^{2-}) - E(CO_2)$$

The results of B3LYP energy of $\Pi (a^{2-}$-$CO_2)$ with different basis sets are list in table 1. In the 2e-reduced states, complex $a^{2-}$ reacted with $CO_2$ to form a π-adducted state. Such reaction can be analogues to $CO_2$ of π reaction with benzene as listed also in Figure 1. Interestingly, the stabilized energy listed in table 1 is ca. -6 Kcal/mol, significant larger than the energy for $CO_2$ with benzene, noted as, -1.3Kcal/mol. In aqueous model (PCM solvation), although the stabilization energy is lower, noted as -3.6 Kcal/mol for former and -0.4 Kcal/mol for later. Such high stabilization energy was due to specific structure orientation of tungsten dithiolene as shown in Figure 1. Only trivial difference of each basis sets were observed, hence 6-31(d) was applied in the further studies (for example, frequency calculations) in order to conduct an economical computation time in reasonable accuracy.

TABLE 1. Stabilized energy of $\Pi$ ($a^{2-}$-$CO_2$), ($a$-$CO_2$), ($a^{2-}$-$HCO_2^-$), ($a$-$HCO_2^-$) under various basis sets for $CO_2$ or formate interaction of $a$ and 2e reduced complex $a^{2-}$. Unit, Kcal/mol

| compd | [(6-311+(d,p) | 6-31(d,p), | 6-31(d) |
|---|---|---|---|
| $a^{2-}$-$CO_2$ | -6.54 | -6.34 | -6.01 |
| $a$-$CO_2$ | -1.94 | -1.84 | -1.81 |
| $a^{2-}$-$HCO_2^-$ | -* | - | - |
| $a$-$HCO_2^-$ | -7.01 | -6.92 | -6.80 |

* no converge can be obtained, $a^{2-}$ and $HCO_2^-$ stay as far as possible due to the same charge.

It is interesting to point out that such π interaction between tungsten dithiolene complex $a$ and $CO_2$ are significantly smaller, noted as -1.8Kcal/mol. Moreo-ver, no converge can be obtained if **a²⁻** reacted with the CO$_2$ reduced product: formate, indicating that the product prefers to leave the catalyst center if CO$_2$ reduction occurred. While a preferred state of complex **a** with formate can be observed as shown in Figure 1 and the stabilized energy noted as, -6.8 Kcal/mol. Inter-conversion between CO$_2$ and formate can be experimentally build up by FDH as electrocatalyst on a graphite carbon electrode, while for the energy level, this computational result demonstrated a perfect energy states: a reduced state of electrocatalyst prefers binding of CO$_2$, and an oxidized state of electrocatalyst prefers binding interaction of formate.

Molecular orbitals of these adduct complexes have also been investigated to directly highlight their π interactions as shown in Figure 2 and Figure 4 respectively. Very little bonding interactions between CO$_2$ and complex **a** is observed in Figure 2. However, such pi-type interaction is quite obvious as shown in Figure 4 for complex **a²⁻** with CO$_2$. It is nicely corroborated with the energy calculation in Table 1 that the reduced states exhibit significantly large stabilization energy. Such bonding condition can be also understood from Figure 3 between CO$_2$ and benzene.

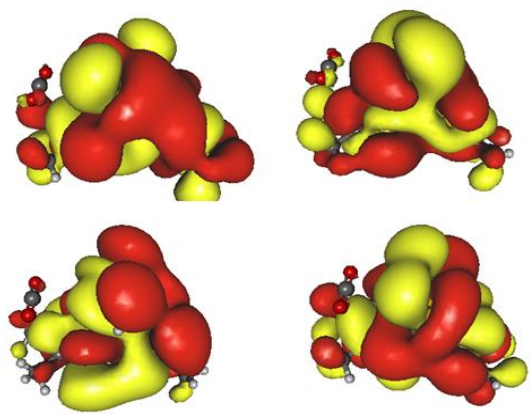

Figure 2. Molecular orbitals (up left, HOMO; up right, HOMO at different iso-surface level for clarity; bottom left, LUMO; bottom right, LUMO+1) of **a-CO2.**

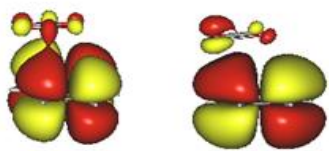

Figure 3. Molecular orbitals of **benzene-CO2.**

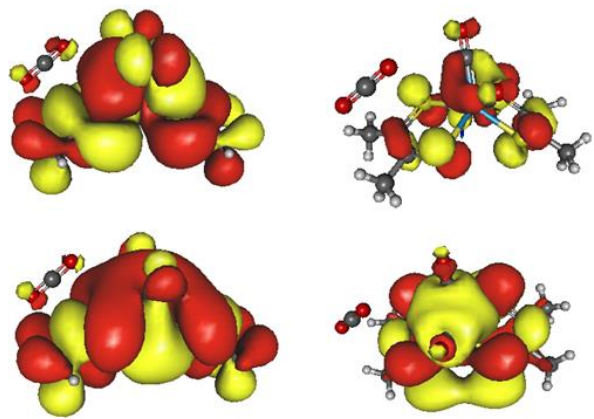

Figure 4. Molecular orbitals (up left, HOMO; up right, HOM-1; bottom left, LUMO; bottom right, LUMO+1) of **a²⁻-CO2.**

The Bailar twisted angle, from 60° to 0°, governing structure preference of tungsten dithiolenes from octahedral to triangle prismatic geometries in their restricted structures, has also been explored to evaluate such π interrelations with CO$_2$ and format. The angle increase was illustrated in Figure 5 and the energy results are listed in Table 2. A restricted octahedral structure at 0° for **a²⁻** with -3 kcal/mol stabilization energy is preferred over the triangle prismatic when such interactions are concerned. Hence it concludes that an octahedral dithiolene complex might be favored over triangle prismatic structure towards CO$_2$ reduction, or at least towards initial binding step. This finding presents a favor of crystal structure of the reported natural FDH enzyme.[11]

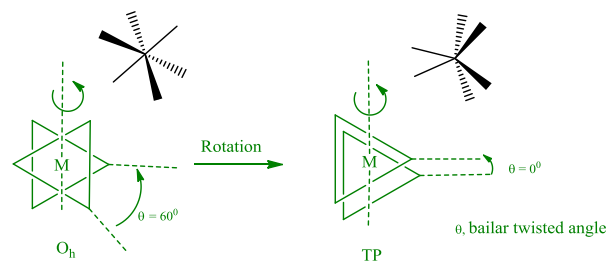

Figure 5. Baior twisted angle for a typical tungsten dithiolene complex.

Table 2. Stabilization energy for a²⁻ with CO2 under restricted Bailar twist angle from 0° to 60°. Unit: Kcal/mol.

| angle | 0 | 10 | 20 | 30 | 40 | 50 | 60 |
|---|---|---|---|---|---|---|---|
| $\Pi_{(a^{2-}-CO_2)}$ | -5.1 | -4.2 | -0.1 | 4.8 | 0.8 | -4.9 | -8.0 |

In conclusion, we have theoretically investigated the tungsten dithiolene complex's π interaction with $CO_2$ and formate. The higher stabilized energy of reduced complex with $CO_2$, and oxidized complex with formate indicated the preferred binding reaction that leads to inter-conversion between $CO_2$ and formate. While little interaction can be observed between $CO_2$ and oxide catalyst and formate with reduced compounds, it provided another favor of aforementioned reaction via product leaving. The molecular orbital analysis also proved and corroborated such π interaction. The Bailar twisted angel incremental studies demonstrated that a restricted octahedral structure with -3 kcal/mol energy is preferred over the triangle prismatic when they are π bonding to $CO_2$. This theoretical investigation might be of great importance to illustrate the preference of octahedral structure of nature dithiolene FDH enzyme's active site.

## ASSOCIATED CONTENT

### Supporting information

Optimized structures, details of methods, molecular orbitals of complex **a** reacting with formate, detailed Bailar twist angle increasing strategy etc.

## AUTHOR INFORMATION

### Corresponding Author

yongchem@gmail.com (YY)

### Present Addresses

†National Renewable Energy Lab, Golden, CO 80401## ACKNOWLEDGMENT

The authors acknowledge center for computational science and technology services at Tulane University for CCS cluster access. We also want to thank Bocarsly group in Princeton University and Donahue Group in Tulane University for helpful discussions.## REFERENCES

(1) (a) Gu, J.; Yan, Y.; Krizan, J. W.; Gibson, Q. D.; Detweiler, Z. M.; Cava, R. J.; Bocarsly, A. B. *J. Am. Chem. Soc.* **2014**, *136*, 830. (b) Walter, M. G.; Warren, E. L.; McKone, J. R.; Boettcher, S. W.; Mi, Q. X.; Santori, E. A.; Lewis, N. S. *Chem Rev* **2010**, *110*, 6446.

(2) (a) Appel, A. M.; Bercaw, J. E.; Bocarsly, A. B.; Dobbek, H.; DuBois, D. L.; Dupuis, M.; Ferry, J. G.; Fujita, E.; Hille, R.; Kenis, P. J. A.; Kerfeld, C. A.; Morris, R. H.; Peden, C. H. F.; Portis, A. R.; Ragsdale, S. W.; Rauchfuss, T. B.; Reek, J. N. H.; Seefeldt, L. C.; Thauer, R. K.; Waldrop, G. L. *Chem. Rev.* **2013**, *113*, 6621. (b) Bocarsly, A.; Gu, J.; Yan, Y.; Watkins, J.; Zeitler, E.; Detweiler, Z.; Hu, Y.; Liao, K.; White, J.; Baruch, M.; Pander, J.; Wuttig, A. *Abstr. Pap. Am. Chem. Soc.* **2013**, *246*. (c) Yan, Y.; Bocarsly, A. *Abstr. Pap. Am. Chem. Soc.* **2013**, *245*. (d) Yan, Y.; Gu, J.; Bocarsly, A. B. *Aerosol Air Qual Res* **2014**, *14*, 515. (e) Yan, Y.; Zeitler, E. L.; Gu, J.; Hu, Y.; Bocarsly, A. B. *J. Am. Chem. Soc.* **2013**, *135*, 14020.

(3) (a) Costentin, C.; Drouet, S.; Robert, M.; Saveant, J. M. *Science* **2012**, *338*, 90. (b) Costentin, C.; Robert, M.; Saveant, J. M. *Chem. Rev.* **2010**, *110*, Pr1. (c) Saveant, J. M. *Chem Rev* **2008**, *108*, 2348.

(4) (a) Inoue, T.; Fujishima, A.; Konishi, S.; Honda, K. *Nature* **1979**, *277*, 637. (b) Kumar, B.; Llorente, M.; Froehlich, J.; Dang, T.; Sathrum, A.; Kubiak, C. P. *Annu. Rev. Phys. Chem.* **2012**, *63*, 541.

(5) Xiang, D. M.; Magana, D.; Dyer, R. B. *J. Am. Chem. Soc.* **2014**, *136*, 14007.

(6) (a) Hine, F. J.; Taylor, A. J.; Garner, C. D. *Coordin Chem. Rev.* **2010**, *254*, 1570. (b) Leopoldini, M.; Chiodo, S. G.; Toscano, M.; Russo, N. *Chem-Eur. J.* **2008**, *14*, 8674. (c) Parkin, A.; Seravalli, J.; Vincent, K. A.; Ragsdale, S. W.; Armstrong, F. A. *J. Am. Chem. Soc.* **2007**, *129*, 10328. (d) Reda, T.; Plugge, C. M.; Abram, N. J.; Hirst, J. *P. Natl. Acad. Sci. USA* **2008**, *105*, 10654.

(7) (a) Greene, A. F.; Chandrasekaran, P.; Yan, Y.; Mague, J. T.; Donahue, J. P. *Inorg. Chem.* **2014**, *53*, 308. (b) Sproules, S.; Banerjee, P.; Weyhermuller, T.; Yan, Y.; Donahue, J. P.; Wieghardt, K. *Inorg. Chem.* **2011**, *50*, 7106. (c) Yan, Y.; Chandrasekaran, P.; Mague, J. T.; DeBeer, S.; Sproules, S.; Donahue, J. P. *Inorg. Chem.* **2012**, *51*, 346. (d) Yan, Y.; Donahue, J. P.; Gu, J.; Mague, J. T. *Abstr. Pap. Am. Chem. Soc.* **2012**, *244*. (e) Yan, Y.; Keating, C.; Chandrasekaran, P.; Jayarathne, U.; Mague, J. T.; DeBeer, S.; Lancaster, K. M.; Sproules, S.; Rubtsov, I. V.; Donahue, J. P. *Inorg. Chem.* **2013**, *52*, 6743. (f) Yan, Y.; Mague, J. T.; Donahue, J. P. *Acta Crystallogr. E.* **2009**, *65*, o1491.

(8) (a) Burgess, B. K.; Lowe, D. J. *Chem. Rev.* **1996**, *96*, 2983. (b) Dobbek, H. *Coordin Chem Rev* **2011**, *255*, 1104 (c) Jormakka, M.; Tornroth, S.; Byrne, B.; Iwata, S. *Science* **2002**, *295*, 1863. (d) Romao, M. J. *Dalton. Trans.* **2009**, 4053.

(9) Hudson, G. A.; Cheng, L.; Yu, J. M.; Yan, Y.; Dyer, D. J.; McCarroll, M. E.; Wang, L. C. *J. Phys. Chem. B* **2010**, *114*, 870.

(10) Gu, J.; Yan, Y.; Helbig, B. J.; Huang, Z.; Lian, T.; H., S. R. *Coord. Chem. Rev.* **2015**, *282-283*, 100.

(11) Schirwitz, K.; Schmidt, A.; Lamzin, V. S. *Protein. Sci.* **2007**, *16*, 1146.